# Next-Level, Robotic Telescope-Based Observing Experiences to Boost STEM Enrollments and Majors on a National Scale

Daniel E. Reichart

*University of North Carolina at Chapel Hill, Department of Physics and Astronomy, Campus Box 3255, Chapel Hill, NC 27599-3255*

**Abstract**. Funded by a $3M Department of Defense (DoD) National Defense Education Program (NDEP) award, we are developing and deploying on a national scale a follow-up curriculum to "Our Place In Space!", or OPIS!, in which ≈3,500 survey-level astronomy students are using our global network of "Skynet" robotic telescopes each year. The goal of this new curriculum, called "Astrophotography of the Multi-Wavelength Universe!", or MWU!, is to boost the number of these students who choose STEM majors. One semester in, our participant program has begun, and participating educators have made good progress on MWU!'s first two modules. Excellent progress has been made on the software front, where we have developed new graphing, analysis, and modeling tools in support of these, and upcoming, modules. On the hardware front, preparation continues to expand Skynet to include a global network of intermediate-sized, radio telescopes, capable of exploring the invisible universe.

## 1. Introduction: Skynet/PROMPT, OPIS!, and ERIRA

Over the past two decades, UNC-Chapel Hill has built one of the largest networks of fully automated, or robotic, telescopes in the world, significantly advancing this new technology. These telescopes are used both for cutting-edge research (see §1.1) and for cutting-edge education (see §1.2): Funded by a series of large NSF awards, (1) we have developed unique, student-level, observing and image-analysis interfaces (see §2.1, §3), allowing students, of all ages, to use this globally distributed, research tool, right alongside the professionals; and (2) often in partnerships with professional educators and education researchers, we have developed a sequence of observation-based curricula and experiences that leverage these hardware and software resources, from the elementary-school level through the graduate-school level, reinforcing and strengthening the STEM pipeline. Funded by a $3M Department of Defense (DoD) National Defense Education Program (NDEP) award, we are now developing a new, primarily undergraduate-level curriculum that will target students as they are deciding their major/minor. This curriculum, called "Astrophotography of the Multi-Wavelength Universe!", or MWU!, will leverage two of our most successful education efforts to date (see §1.2, §1.3), with the end goal of significantly boosting STEM enrollments and STEM majors and minors on a national scale, as well as boosting students' technical and research skills.

### 1.1. Skynet and PROMPT

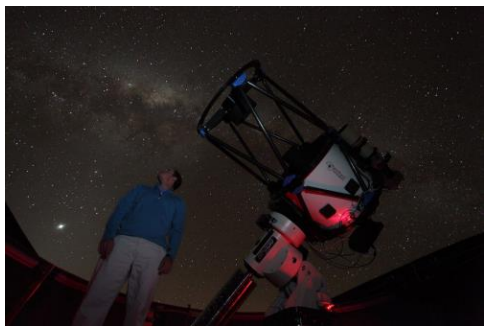

Figure 1. One of our 24-inch diameter robotic telescopes, at Cerro Tololo Inter-American Observatory (CTIO) in the Chilean Andes.

UNC-Chapel Hill began building "Skynet" and "PROMPT" in 2004. Skynet is sophisticated telescope control and queue scheduling software, that provides capacity to be a novel educational technology. Skynet can control scores of telescopes simultaneously, including most commercially available telescope hardware, and provides participating institutions with easy-to-use web (see §2.1, §3) and API interfaces. Participating institutions are not charged, but instead contribute 10% of each of their telescopes' time for Director Discretionary science (e.g., gravitational-wave events[1]) and education (see §1.2). The Skynet Robotic Telescope Network (RTN) has grown to number ≈20 telescopes, with another ≈15 scheduled for integration during the grant period. Skynet's optical telescopes range in size from 14 to 40 inches, and span four continents and five countries. Skynet also includes a 20m-diameter radio telescope (see Figure 3) [5], with plans to integrate 6 – 8 more in support of MWU! (see §2.2).

PROMPT is a subset of the Skynet RTN, consisting of our highest-quality optical telescopes at our highest-quality sites. Originally only at CTIO (Figure 1), PROMPT now spans four and soon five dark sites, in Chile, Australia (for near-continuous observing in the southern hemisphere), and Canada (for full-sky coverage).

Skynet data are now published in peer-reviewed journals every ≈20 days, including five times to date in Nature and Science [3,6-9]. Furthermore, these publications (≈170 refereed, ≈720 total) have been of increasingly high impact, so far resulting in ≈8,600 citations.

**1.2. Skynet-based Curricula and OPIS!**

To date, ≈45,000 students have used Skynet, with most now participating in three large, NSF-funded programs: (1) Skynet Junior Scholars (SJS, $1.6M), which is being carried out in partnership with 4H, for middle-school-age students; (2) Innovators Developing Accessible Tools for Astronomy (IDATA, $2.5M), for high-school students exploring computational thinking in astronomy and helping to develop image-analysis software for blind and visually impaired users; and (3) Our Place In Space! (OPIS!, $2.1M), a Skynet-based laboratory curriculum for survey-level undergraduate and advanced high-school students [10].

OPIS! is a sequence of eight, and soon nine, labs in which students use the same research instrumentation as professionals to collect their own data. They then use this self-collected data (astronomical images and spectra) to reproduce some of the great-est astronomical discoveries of the past 400 years, and gain technical and research skills at the same time. Although students are not carrying out cutting-edge research, they are using cutting-edge research instrumentation, and consequently there is great overlap with the Course-based Undergraduate Research Experience (CURE) pathway model [e.g., 11]. And these labs/observing experiences are specifically designed to pair with standard introductory astronomy curricula, facilitating widespread adoption.

---

[1] Skynet co-discovered the first, and so far only, optical counterpart to a gravitational-wave event in 2017 [1-4].

In UNC-Chapel Hill's first five years using OPIS!, introductory astronomy enrollments increased by >100% – now one in six UNC-Chapel Hill students take at least one of our introductory astronomy courses – and astronomy-track majors and minors increased by ≈300% (from ≈10 to ≈40; however, see below). OPIS! has since been adopted by ≈2 dozen institutions nationally, with another ≈1 dozen adopting next year, and recent adopters are reporting similar enrollment gains. OPIS! currently serves ≈3,500 students per year, and we have sufficient telescope access to increase this tenfold or more over the next decade.

However, while OPIS! has succeeded at increasing STEM enrollments at UNC-Chapel Hill, and is now being scaled to do so nationally, we cannot claim that it is solely responsible for our dramatic increase in majors and minors: OPIS! has significantly increased our number of students who return to take another, "next-level" astronomy course, but it is a weeklong field experience that we expose many of these, returning students to that "seals the deal". We describe this experience in the next section, after which we describe our new, DoD-funded effort to develop a next-level, OPIS!-style curriculum that adapts it to the classroom (in person or online), and in such a way that it too can scale to a national level.

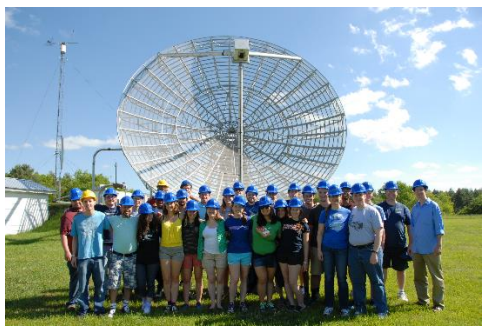

Figure 2. GBO's fully manual, 12m-diameter, teaching radio telescope, and ERIRA participants and educators.

### 1.3. ERIRA

"Educational Research in Radio Astronomy" is a unique, introductory research and field experience in astronomy. For one week each summer since 1992, a small group of volunteer, radio-astronomy educators from across the country have been taking 18 primarily first- and second-year undergraduate students (who have taken at least one astronomy course, but who have not yet decided upon their major) to Green Bank Observatory (GBO) in WV for a 24/7 crash course in radio astronomy.

Radio observing is a powerful teaching tool. Logistically, it can be done during the day when students are naturally awake, and through most weather conditions. In combination with optical observing, it can foster a deeper understanding of the electromagnetic spectrum, and of the important role that multi-wavelength observations play in 21st-century astronomy: Students can be exposed to a wide variety of astrophysical phenomena – solar-system objects, star-forming regions, supernova remnants and pulsars, galaxies, etc. – and emission processes – blackbody, synchrotron, bremsstrahlung, and a variety of emission-line mechanisms – in ways that are fundamentally different from if they are experienced in only one waveband or the other.

*4*        ReichartUsing GBO's 12m (Figure 2), as well as GBO's 20m (see Figure 3) through Skynet, and Skynet optical telescopes around the world (e.g., Figure 1), the students carry out nearly a dozen "next-level", multi-wavelength observing experiences. ≈85% of ERIRA participants have become physics and astronomy majors and minors upon their return; ≈100% have become STEM majors and minors.

## 2. MWU!: Next-Level, Multi-Wavelength Observing Experiences

Funded by DoD, we are now: (1) developing a collection of eight new, multi-wavelength observing experiences called "Astrophotography of the Multi-Wavelength Universe!" (MWU!), based largely on ERIRA, but instead making use of Skynet so they can be incorporated into these next-level (but still pre-major/minor decision) courses and serve orders of magnitude more students than a field experience can; and (2) through an instructor participant program spanning ≈30, adopting institutions, (a) begin to scale it to a national audience, and (b) simultaneously carry out a deep evaluation of its effectiveness. This effort has multiple components:

### 2.1. Software

MWU! will leverage two significant software development efforts that we completed for OPIS!, and for our middle- and high-school efforts (SJS, IDATA; §1.2). The first is our web-based, observation-request interface, which is used by professionals and students alike: https://skynet.unc.edu.[2] Skynet's backend then acquires the requested images from the requested telescopes, and returns them to the user via the same interface.

However, acquiring images is only half the battle: Students then need to be able to make fundamental measurements from their images. As such, we have additionally developed Afterglow Access (AgA): https://afterglow.skynet.unc.edu.[3] Currently, AgA is used to analyze monochromatic, optical images. As part of this effort, we are significantly expanding its capabilities, to include the analysis of: (1) color optical images, incorporating key aspects of astrophotography as an additional hook (e.g., see Figure 4, left panel); (2) radio images, for which we have already developed and extensively tested new, cutting-edge algorithms [5,12], capable of automatically removing increasingly troublesome, manmade sources of radio interference (e.g., see Figure 4, right panel); and (3) spectra, from either optical or radio telescopes. AgA is a long-desired technology in astronomy education.

### 2.2. Hardware

Skynet includes ≈20 optical telescopes, but so far only one radio telescope (see Figure 3). These are complicated, expensive instruments that even most professionals do not have access to. As with optical telescopes: (1) Each views a different part of the sky, depending on its latitude; (2) Each has different capabilities, e.g., some have lower-frequency receivers necessary for detecting cold, hydrogen gas, others have higher-frequency receivers necessary for spatially resolving structures (e.g., see Figure 4, right panel); and (3) Each may go offline for days, and sometimes even weeks, because of required maintenance. As such, a globally distributed network, featuring both diversity and redundancy, is highly desirable.

---

[2] For a video tutorial: https://tinyurl.com/skynet-tutorial
[3] For multiple video tutorials: https://tinyurl.com/afterglow-tutorial

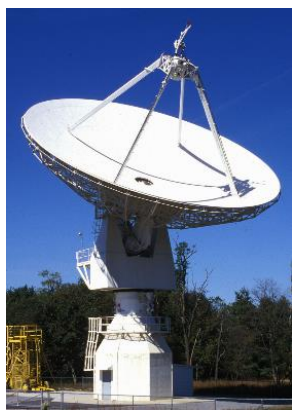

Figure 3. Skynet's 20m-diameter radio telescope at GBO, which is used by thousands of students per year via Skynet.[4]

We will soon be integrating: (1) Arecibo Observatory's 12m telescope in Puerto Rico, (2) 4.5m, 12m, and (contingent upon board approval) two 26m telescopes at the Pisgah Astronomical Research Institute (PARI) in North Carolina; and (3) 14m, 26m, and 30m telescopes at two locations in Australia, under the management of the University of Tasmania, giving students southern-hemisphere access as well. This will also require some generalization of Skynet's backend software.

This is the first time that multiple, diverse radio telescopes will be broadly, and easily, accessible through a single interface, and will likely establish a standard that other radio telescopes will adopt, permitting their unused/background-job observing time to be used by students. The impact of this will likely be far-reaching, beyond that of the MWU! curriculum.

**2.3. Curriculum**

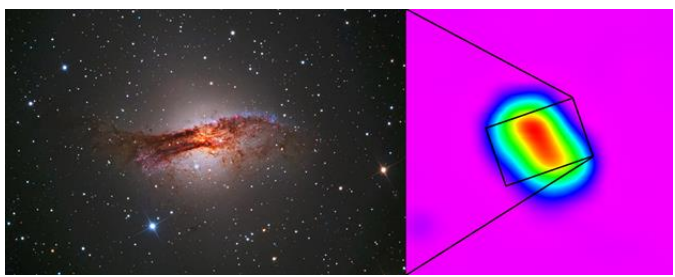

Figure 4. PROMPT color image (left) and GBO 20m 9-GHz image (right) of Centaurus A, which will be the focus of one of MWU!'s two capstone observing experiences, integrating both optical and radio observations and analyses. Consisting of an elliptical galaxy that recently merged with a smaller spiral galaxy, the influx of new gas resulted in a burst of star formation. Students will image the galaxy using red, green, blue, and narrowband filters, probing new vs. old star formation, and dust lanes (left), and will collect 1.4-GHz radio spectra, probing cold, hydrogen gas. Centaurus A also features a 55-million-solar-mass black hole, which is now active, producing galaxy-scale jets, which students can (at least partially) resolve by imaging at higher radio frequencies (right).

---

[4] For a video tutorial: https://tinyurl.com/radio-skynet-tutorial

MWU! will consist of eight new observing experiences (three optical, three radio, and two capstone experiences involving both; e.g., Figure 4) based on ERIRA's manual observing experiences. These observing experiences will strongly reinforce the subject matter of next-level astronomy courses, which are typically on stars, galaxies, and their multi-wavelength emission (and absorption) mechanisms. As with OPIS!, MWU! will work equally well online as in person – Skynet and AgA require only an internet connection. Indeed, OPIS! was one of the few STEM labs that transitioned seamlessly when the pandemic began.

## 3. MWU!: First-Semester Progress

### 3.1. Curriculum

After a few organizational meetings, our participant program began on September 30, 2021, 15 days after the start of the award. Currently 26 educators are participating; most are faculty members at the OPIS! institutions. The educators are divided into two groups, and each group is meeting weekly. For each of the first two years, each group is responsible for producing two of the eight modules of the MWU! curriculum, and trialing them with their students. We are currently tackling the first and fourth modules of the curriculum:

*Module 1:*  In this module, students will break into teams and each team will use Skynet to acquire images of one or more star clusters, at three or four different optical wavelengths. From these images, they will photometer all stars in their images (typically a few thousand), and calibrate these values (see §3.2). They will then use these data to plot HR diagrams, and fit isochrone models to these data to date the cluster (see Figure 5). Lastly, they will combine their multi-wavelength images into a single, deep color picture (see §3.2), and develop an intuitive understanding for the relationship between ages and colors of stellar populations.

To date, the educators have carried out the first half of this activity (through making HR diagrams) for a variety of star clusters, and have compiled a list of suitable star clusters that the module can be designed around. Some have begun working through this list with their students, and a first draft observation and analysis guide has been fleshed out. Work has also been completed on the module's background sections, with a general outline agreed to by all educators, and drafting work beginning.

*Module 4:*  In this module, students will use Skynet to observe a variety of bright, slow pulsars at radio wavelengths, to introduce them to non-thermal light-emitting mechanisms. Some of the pulsars will be sufficiently bright to see individual pulses (see Figure 6); others will be lost in the noise and extracted by identifying the pulsar's period (see Figure 7), and then folding the data at this period, to beat down the noise (see Figure 8). By calibrating and differencing orthogonal polarization channels in this period-folded data, students show the emission to be polarized, and hence non-thermal.

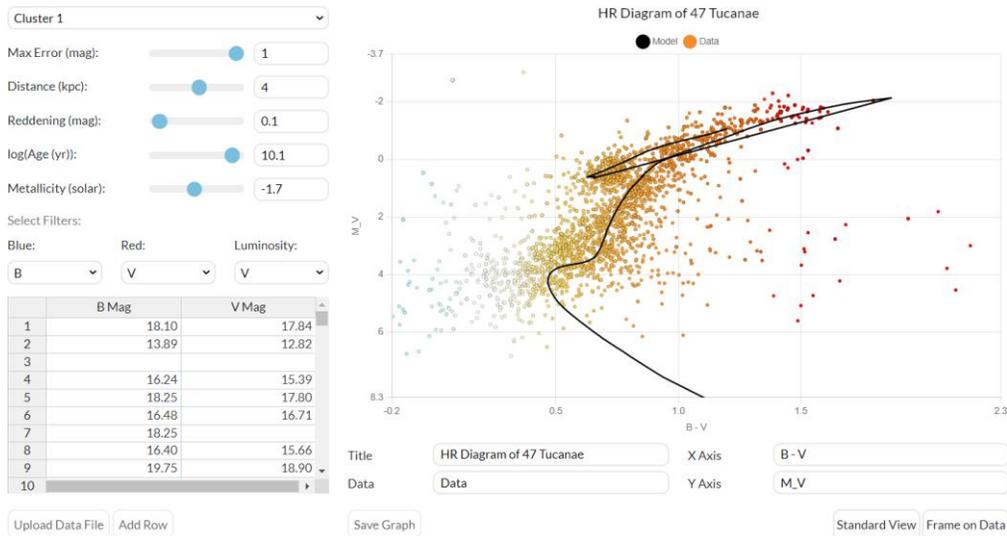

Figure 5. HR diagram of 47 Tucanae, which isochrone modeling reveals it to be over 10 billion years. These data were collected with one of Skynet's optical telescopes at CTIO (Figure 1).

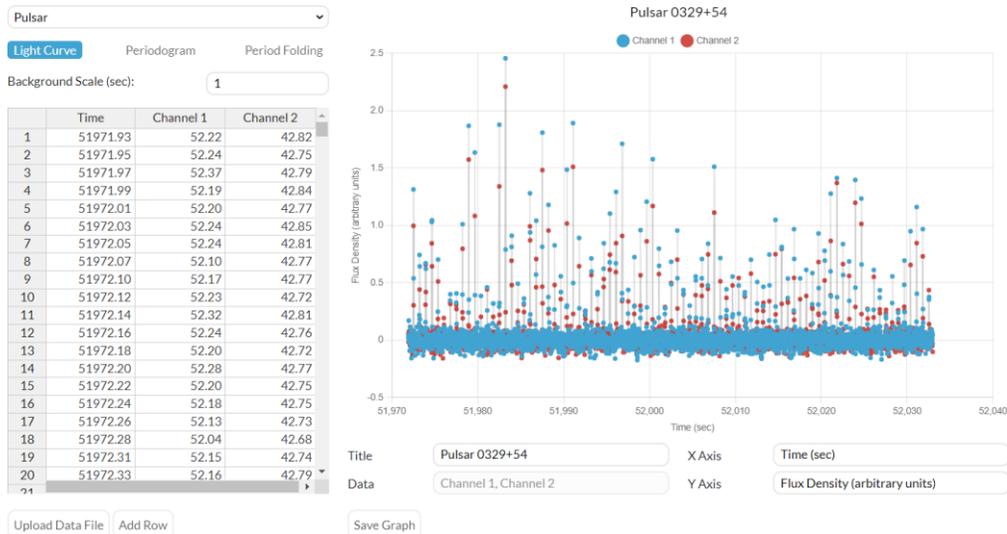

Figure 6. One minute of data from bright, slow pulsar 0329+54, collected with Skynet's 20-meter diameter telescope at GBO (Figure 3).

To date, the educators have carried out this activity for a variety of pulsars, using the one radio telescope that we have already integrated into Skynet (§2.2). They have compiled a list of suitable pulsars that this module can be designed around, and some have begun working through this list with their students. The educators have agreed on a 3-part structure for the module, where some institutions might do only 1 or 2 parts, but most will all 3 parts. Drafting work is beginning.

8             Reichart8             Reichart

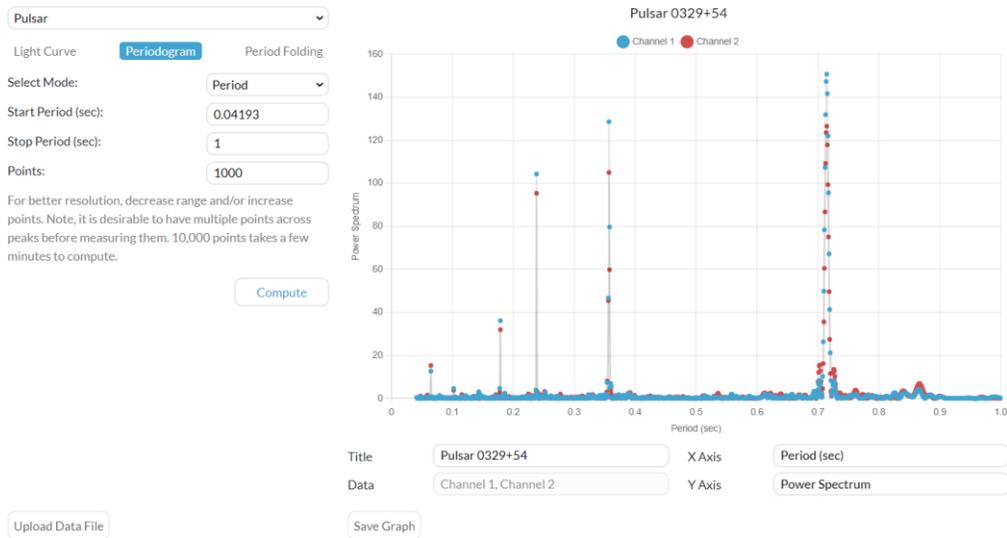

Figure 7. Periodogram of the data from Figure 6, computed using Lomb-Scargle (similar to a Fourier transform, but for non-uniformly sampled data). This pulsar is rotating every 0.7145 seconds.

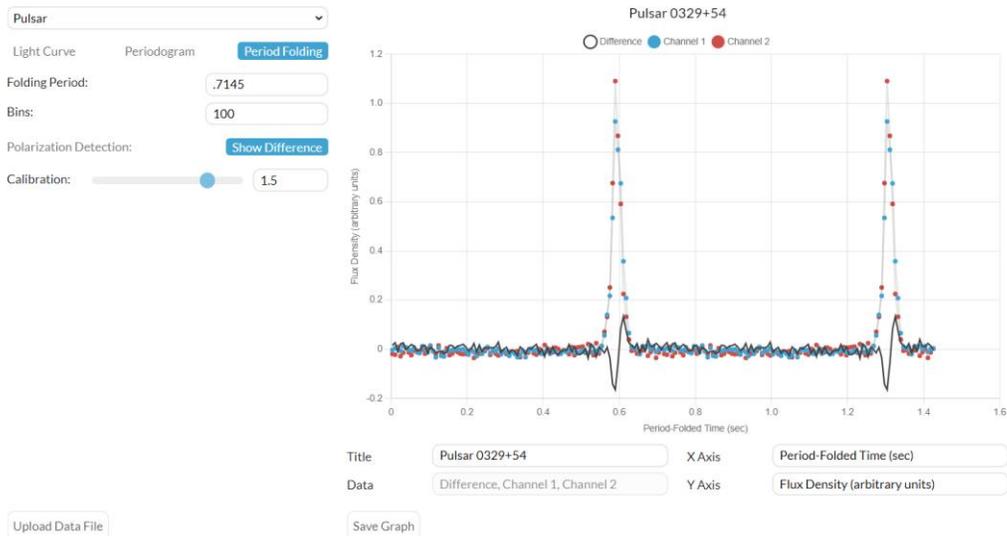

Figure 8. Period-folding of the data from Figure 6. The non-zero difference in the orthogonal polarizations channels shows that one channel's pulse leads the other's, demonstrating polarization and non-thermal emission (caused by electrons moving along the rotating neutron star's magnetic field lines).

### 3.2. Software

The most immediate software need this semester was developing the graphing/modeling/analysis interfaces that the educators needed to work on Modules 1 and 4, and that will eventually be used by their students. We decided to add these to the collection of interfaces that we have already developed for OPIS!: https://tinyurl.com/skynet-graph (Figures 5-8).

While these tools were developed by undergraduate programmers, Skynet's two full-time, professional programmers pushed on other fronts. First, they modified AgA, our online image processing and analysis application (§2.1), so students can not only carry out photometry, but calibrated photometry, of thousands of stars per image, for many images simultaneously (see Figure 9). This was a significant effort, without which the Module 1 cluster interface described above would be useless. Students using AgA can now tie their photometric measurements to a wide selection of modern catalogs, allowing them to not only complete MWU! exercises, but to transition to publishable research using the exact same application. As part of this effort, we added professional-level photometry and source-extraction algorithms as well.

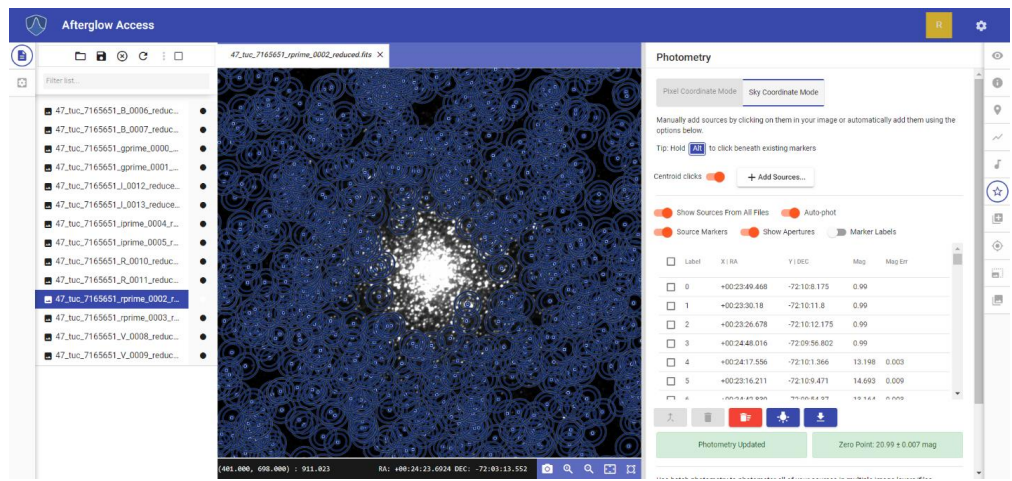

Figure 9. Screenshot of AgA's expanded photometry tool, applied to 47 Tucanae, the source of the data in Figure 5.

In preparation for the second part of Module 1, and for all of the upcoming MWU! modules, Skynet's full-time programmers also added color combination capabilities to AgA (see Figure 10). Additional features are being added to this now, some piggybacking off of the above, new photometric calibration capability, which will allow students to balance the red, green, and blue channels as they are in nature.

**4. Additional Information**

For additional information about OPIS! and MWU!, and implementing them at your institution: (1) https://arxiv.org/abs/2103.09895 (OPIS!), (2) https://tinyurl.com/skynet-workshop, (3) https://tinyurl.com/skynet-links, and/or (4) email introastro@unc.edu.

**Acknowledgements**. We grateful acknowledge the Department of Defense, for its support through National Defense Education Program (NDEP) award

HQ00342110018. We also gratefully acknowledge the support of the National Science Foundation, through the following programs and awards: ESP 0943305, ISE 1223235, HBCU-UP 1238809, TUES 1245383, STEM+C 1640131, AAG 2007853, and IUSE 2013300. Lastly, we thank the Mt. Cuba Astronomical Foundation, the North Carolina Space Grant Consortium, and the University of North Carolina System for their support.

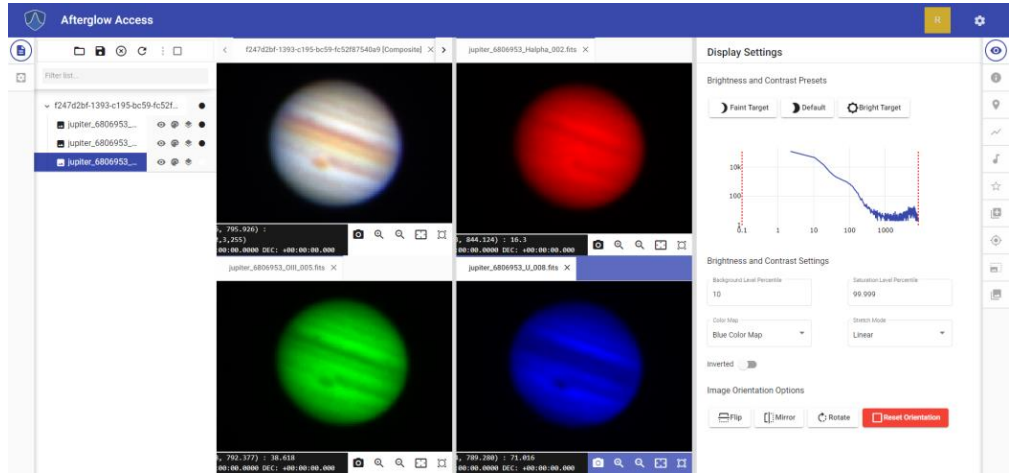

Figure 10. Screenshot of AgA's color combination tool, applied to 0.8 seconds of imaging data of Jupiter.